\documentclass[%
 reprint,
superscriptaddress,
 amsmath,amssymb,
 aps,
prl,
]{revtex4-2}

\usepackage{graphicx}
\usepackage{dcolumn}
\usepackage{bm}
\usepackage{braket}
\usepackage{comment} 
\usepackage{hyperref}
\usepackage{siunitx}
\usepackage[dvipsnames]{xcolor}
\definecolor{mycitationcolor}{RGB}{91,68,214}
\hypersetup{pdfborder=0 0 0} 
\hypersetup{
    colorlinks=true,
    citecolor=Blue,
    linkcolor=Blue,
    urlcolor=Blue
}

\usepackage{balance}

\usepackage{nicematrix}
\NiceMatrixOptions{cell-space-limits = 1pt}

\usepackage{mathtools}

\usepackage{amsthm}

\newtheorem*{theorem*}{Theorem}

\newtheorem*{proposition}{Proposition}

\theoremstyle{definition}

\theoremstyle{remark}

\newtheorem*{example*}{Example}
\newtheorem*{algorithm*}{Algorithms}

\begin{document}

\title{Probability Distribution for Coherent Transport of Random Waves}

\author{Yunrui Wang}
\affiliation{Department of Electrical and Computer Engineering and Microelectronics Research Center, The University of Texas at Austin, Austin, Texas 78712, USA}

\author{Cheng Guo}
\email{chengguo@utexas.edu}
\affiliation{Department of Electrical and Computer Engineering and Microelectronics Research Center, The University of Texas at Austin, Austin, Texas 78712, USA}

\date{\today}

\begin{abstract}
We establish a comprehensive probability theory for coherent transport of random waves through arbitrary linear media. The transmissivity distribution for random coherent waves is a fundamental B-spline with knots at the transmission eigenvalues. We analyze the distribution's shape, bounds, moments, and asymptotic behaviors. In the large $n$ limit, the distribution converges to a Gaussian whose mean and variance depend solely on those of the eigenvalues. This result resolves the apparent paradox between bimodal eigenvalue distribution and unimodal transmissivity distribution.
\end{abstract}
\maketitle

\begin{figure}[htbp]
\centering
\includegraphics[width = 0.45\textwidth]{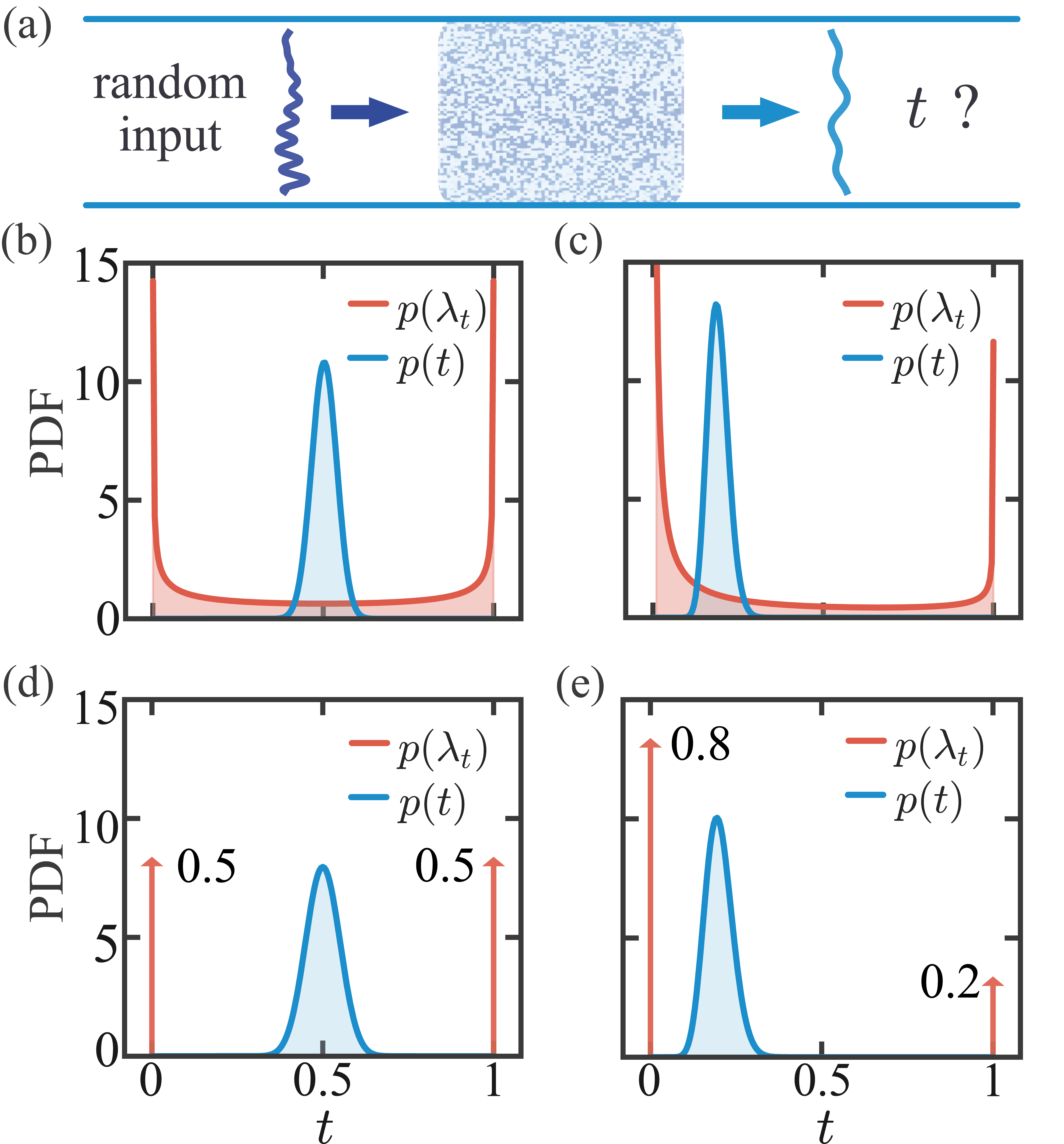}\caption{(a) Central problem: What is the probability distribution $p(t)$ of transmissivity $t$ for random coherent waves incident on a scattering medium? (b,c) Transmission eigenvalue distribution $p(\lambda_t)$ (red) versus Monte Carlo-simulated transmissivity distribution $p(t)$ (blue) for (b) fully chaotic and (c) diffusive systems. (d,e) Corresponding results for Bernoulli-distributed eigenvalues with (d) $q=0.5$ and (e) $q=0.8$.}
\label{fig:intro}
\end{figure}

\textit{Introduction}---Wave transport in complex media is a fundamental problem in physics~\cite{lee1985b,datta1995,sheng2006,taylor2006,akkermans2011a,nazarov2012,zhang2023g}, giving rise to phenomena such as Anderson localization~\cite{andersonAbsenceDiffusionCertain1958,lee1985disordered,evers2008anderson}, coherent backscattering~\cite{albadaObservationWeakLocalization1985,wolf1985weak}, random lasing~\cite{cao1999random,caoReviewLatestDevelopments2005,wiersma2008physics}, and coherent perfect absorption~\cite{chongCoherentPerfectAbsorbers2010,wanTimeReversedLasingInterferometric2011,sunExperimentalDemonstrationCoherent2014,baranovCoherentPerfectAbsorbers2017,sweeneyPerfectlyAbsorbingExceptional2019,wangCoherentPerfectAbsorption2021,guoSingularTopologyScattering2023}. The complexity of these systems necessitates a statistical approach to characterize universal wave behaviors [Fig.~\ref{fig:intro}(a)]. Random matrix theory provides such a framework through the statistics of transmission eigenvalues~\cite{wignerStatisticalDistributionWidths1951,wignerCharacteristicVectorsBordered1955}. A key prediction is the existence of open/closed transmission eigenchannels with near-unity/zero eigenvalues~\cite{dorokhov1984,rotter2017light,cao2022shaping}. For chaotic systems with many channels, the transmission eigenvalues follow a bimodal distribution: $p(\lambda_t) = 1/[\pi\sqrt{\lambda_t(1-\lambda_t)}]$ [Fig.~\ref{fig:intro}(b)]~\cite{baranger1994,jalabert1994}. For diffusive systems, the distribution becomes asymmetric: $p(\lambda_t) \propto 1/[\lambda_t\sqrt{1-\lambda_t}]$ [Fig.~\ref{fig:intro}(c)]~\cite{dorokhov1984,imry1986,pendry1997}. Both distributions peak near $0$ and $1$, suggesting that open and closed channels should be readily observable through wavefront shaping~\cite{vellekoop2008universal,popoff2010,choi2011,kim2012a,popoff2014,kim2015c,mastiani2022wavefront,cheng2023high,alhulaymi2025}.

Despite these predictions, experimental observation of open channels has been a significant challenge~\cite{goetschy2013filtering,gerardin2014full,meer2021,lin2024b}. This difficulty has been attributed to incomplete mode access---missing even a single mode can hide open channels entirely~\cite{yu2013a,goetschy2013filtering,meer2021}. Such extreme sensitivity is unexpected and suggests a fundamental gap in our understanding. This observation prompts us to examine an implicit presumption of random matrix theory: that transmission statistics through random media are directly characterized by the eigenvalue distribution~\cite{beenakker1997random}. To test this premise, we performed Monte Carlo simulations of $10^5$ random input waves through both chaotic and diffusive systems with 100 ports. The blue curves in Figs.~\ref{fig:intro}(b) and~\ref{fig:intro}(c) show the resulting transmissivity distributions $p(t)$. Strikingly, while the eigenvalue distributions are bimodal, the transmissivity of random waves follows a unimodal Gaussian-like distribution, with negligible probability of accessing open or closed channels. We further considered extreme cases where eigenvalues follow a Bernoulli distribution~\cite{bertsekas2008}---either fully closed (probability $q$) or fully open (probability $1-q$). Even here, $p(t)$ remains Gaussian for both $q=0.5$ and $q=0.8$. This stark contrast between $p(t)$ and $p(\lambda_t)$ calls for a comprehensive theory of the transmissivity distribution $p(t)$.

In this paper, we establish a comprehensive probability theory for coherent transport of random waves. We prove that the transmissivity distribution $p(t)$ for random waves through any medium is a fundamental B-spline with knots at the transmission eigenvalues. We analyze the distribution's properties---shape, bounds, and moments---and examine its asymptotic behavior for large port numbers. We show that $p(t)$ converges to a Gaussian whose mean and variance depend solely on those of the eigenvalues; all other details of the eigenvalue distribution become irrelevant. We extend this framework to other transport observables, including reflection and absorption. Our results provide a rigorous foundation for understanding coherent wave transport statistics and clarify the precise role of transmission eigenvalues.

\begin{figure}[htbp]
\centering
\includegraphics[width = 0.45\textwidth]{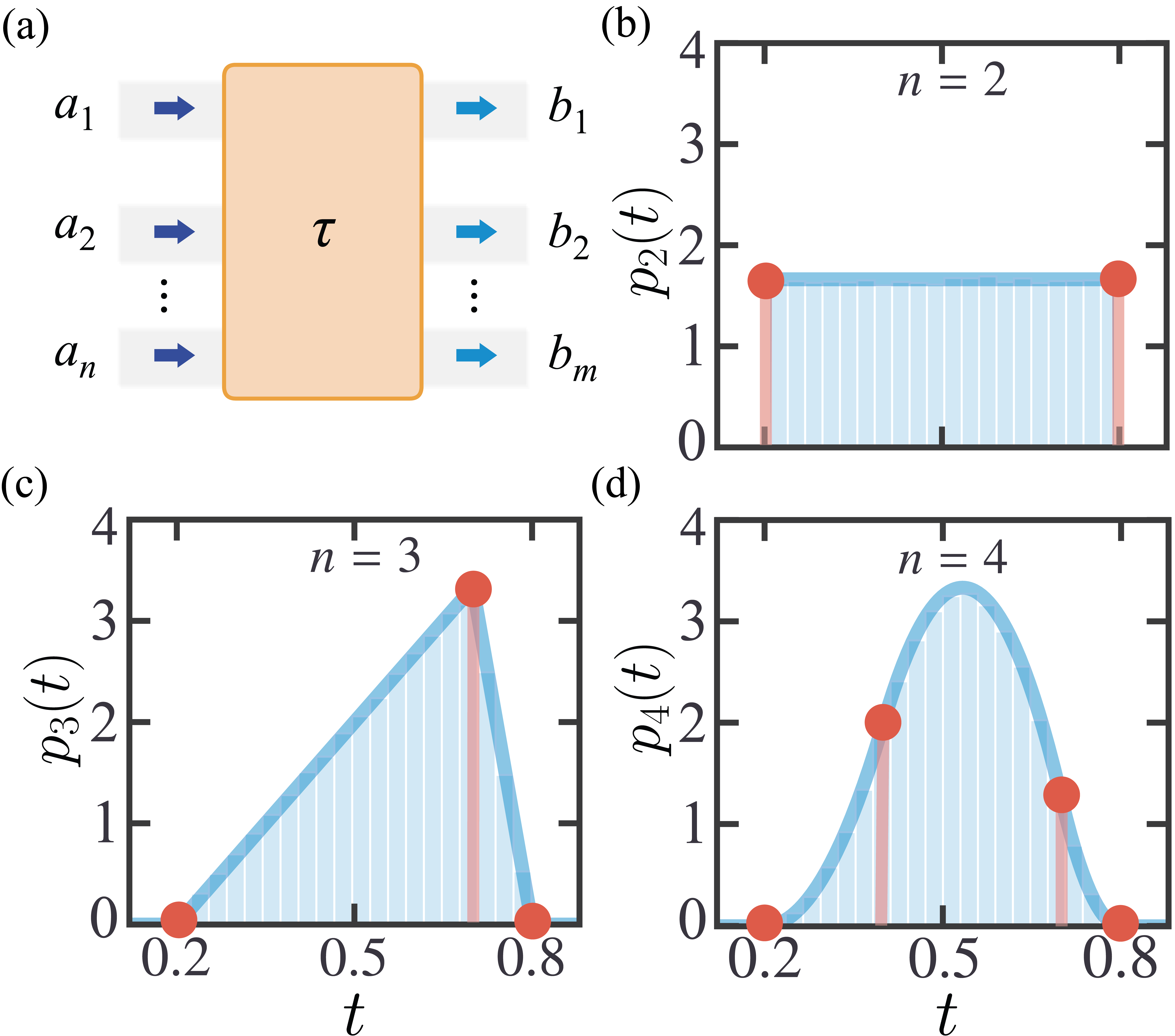}
\caption{(a) An $(n+m)$-port linear system transforms a coherent input wave $\bm{a}$ into a transmitted wave $\bm{b} = \tau \bm{a}$. (b-d) Transmissivity distribution $p_n(t)$ for $n=2, 3, 4$ input ports. Histograms show Monte Carlo results from $10^5$ random coherent inputs, blue curves show analytical B-spline predictions, and red dots and lines mark transmission eigenvalues $\bm{\lambda}(\tau^\dagger \tau)$.}
\label{fig:system}
\end{figure}

\textit{Problem}---Consider an $(n+m)$-port linear time-invariant system with $n$ ports on the left and $m$ ports on the right, as illustrated in Fig.~\ref{fig:system}(a). The system is characterized by the block scattering matrix~\cite{beenakker1997random,rotter2017light}
\begin{equation}\label{eq:block-S-matrix}
    S = \begin{pmatrix}
        \rho & \tau' \\
        \tau & \rho' 
    \end{pmatrix},
\end{equation}
where $\rho$ ($\rho'$) and $\tau$ ($\tau'$) denote the field reflection and transmission matrices for waves incident from the left (right), respectively. We focus on coherent wave transmission from left to right. A normalized coherent input wave, represented by a complex unit vector $\bm{a} = (a_1, \ldots, a_n)^T$, produces a transmitted wave $\bm{b} = \tau \bm{a}$. The corresponding transmissivity is given by
\begin{equation}\label{eq:def_transmissivity}
    t[\bm{a}] \coloneqq \bm{a}^\dagger \tau^\dagger \tau \bm{a} = \bm{a}^\dagger T \bm{a}, 
\end{equation}
where $T = \tau^\dagger \tau$ is the transmittance matrix~\cite{guo2025}, whose eigenvalues are the transmission eigenvalues. We study the transmissivity $t[\bm{a}]$ when $\bm{a}$ is a random coherent input, where ``random'' means that $\bm{a}$ is drawn uniformly from the complex unit sphere $\mathcal{S}^{2n-1} \coloneqq \{\mathbf{z}\in\mathbb{C}^n : |\mathbf{z}|=1\}$. This reflects the absence of any \emph{a priori} preference among input waves. Our aim is to determine the probability density function (PDF)~\cite{bertsekas2008} of the transmissivity:
\begin{equation}
    p_{n}(t) = p_n(t; T).
\end{equation}
We omit the parameter $T$ when no ambiguity arises.

\textit{Solution}---To motivate the general solution, we begin with numerical experiments for small $n$ before examining the large-$n$ limit. Consider a $2 \times 2$ transmittance matrix 
\begin{equation}
    T_2 = \begin{pmatrix}
    0.58 & 0.25+0.14j\\
    0.25-0.14j & 0.42
    \end{pmatrix}, \quad  \bm{\lambda}(T_2) = (0.2, 0.8),
\end{equation}
where $\bm{\lambda}(\cdot)$ denotes the vector of eigenvalues in nondecreasing order for a Hermitian matrix. We sample $10^5$ random $\bm{a}$ from $\mathcal{S}^{2n-1}$ and compute $t[\bm{a}]$ using Eq.~(\ref{eq:def_transmissivity}). Figure~\ref{fig:system}(b) plots the resulting histogram. We observe that $p_2(t;T_2)$ is uniform over the interval $[0.2,0.8]$.

We then perform similar analyses for a $3 \times 3$ transmittance matrix $T_3$ and a $4 \times 4$ transmittance matrix $T_4$ [see~Supplementary Material (SM) \hyperref[em:1]{I} for explicit expressions] with 
\begin{equation}
\bm{\lambda}(T_3) = (0.2, 0.7, 0.8), \quad \bm{\lambda}(T_4) = (0.2, 0.4, 0.7, 0.8).
\end{equation}
Figures~\ref{fig:system}(c) and (d) plot the resulting histograms for $T_3$ and $T_4$, respectively. The distribution $p_3(t;T_3)$ exhibits a triangular shape with vertices at $\bm{\lambda}(T_3)$, while $p_4(t;T_4)$ forms a piecewise quadratic function with continuous derivative, where the knots (subinterval endpoints) coincide with the eigenvalues $\bm{\lambda}(T_4)$.

These numerical results suggest the following characterization: For an $n\times n$ transmittance matrix $T$ with nondegenerate eigenvalues $\lambda_1 < \lambda_2 < \cdots < \lambda_n$, the distribution $p_n(t; T)$ possesses four key properties: (i) it vanishes for $t<\lambda_1$ and $t>\lambda_n$; (ii) it is a polynomial of degree $n-2$ on each subinterval $[\lambda_k, \lambda_{k+1}]$; (iii) it exhibits $n-3$ continuous derivatives at each eigenvalue; and (iv) $\int_{-\infty}^{\infty} p_n(t) \,\mathrm{d}t = 1$. These conditions uniquely determine a function known as the fundamental B-spline, denoted by $M_{n-1}[t;\bm{\lambda}(T)] = M_{n-1}(t;\lambda_1,\ldots, \lambda_n)$~\cite{curry1966,olshanski1996,dunkl2011,gallay2012a}. This observation leads to the central result of this paper:
\begin{equation}\label{eq:main_result}
    p_n(t;T) = M_{n-1}[t;\bm{\lambda}(T)].
\end{equation}
The explicit form of Eq.~(\ref{eq:main_result}) is~\cite{curry1966,deboor1972,deboor2001,schumaker2010}
\begin{equation}\label{eq:main_result_explicit}
    p_n(t;T) = (n-1) \sum_{k=1}^{n} \frac{[\max(\lambda_k - t, 0)]^{n-2}}{\prod_{i \neq k} (\lambda_k - \lambda_i)}. 
\end{equation}
See~\hyperref[em:2]{SM II} for a detailed proof of Eq.~(\ref{eq:main_result}).

We now discuss the general properties of the probability density function $p_n(t;T)$. Equation~(\ref{eq:main_result}) shows that $p_n(t;T)$ is a B-spline completely determined by the transmission eigenvalues $\bm{\lambda}(T)$. Our discussion therefore focuses on how $\bm{\lambda}(T)$ controls the properties of $p_n(t;T)$.

\begin{figure}[htbp]
\centering
\includegraphics[width = 0.28\textwidth]{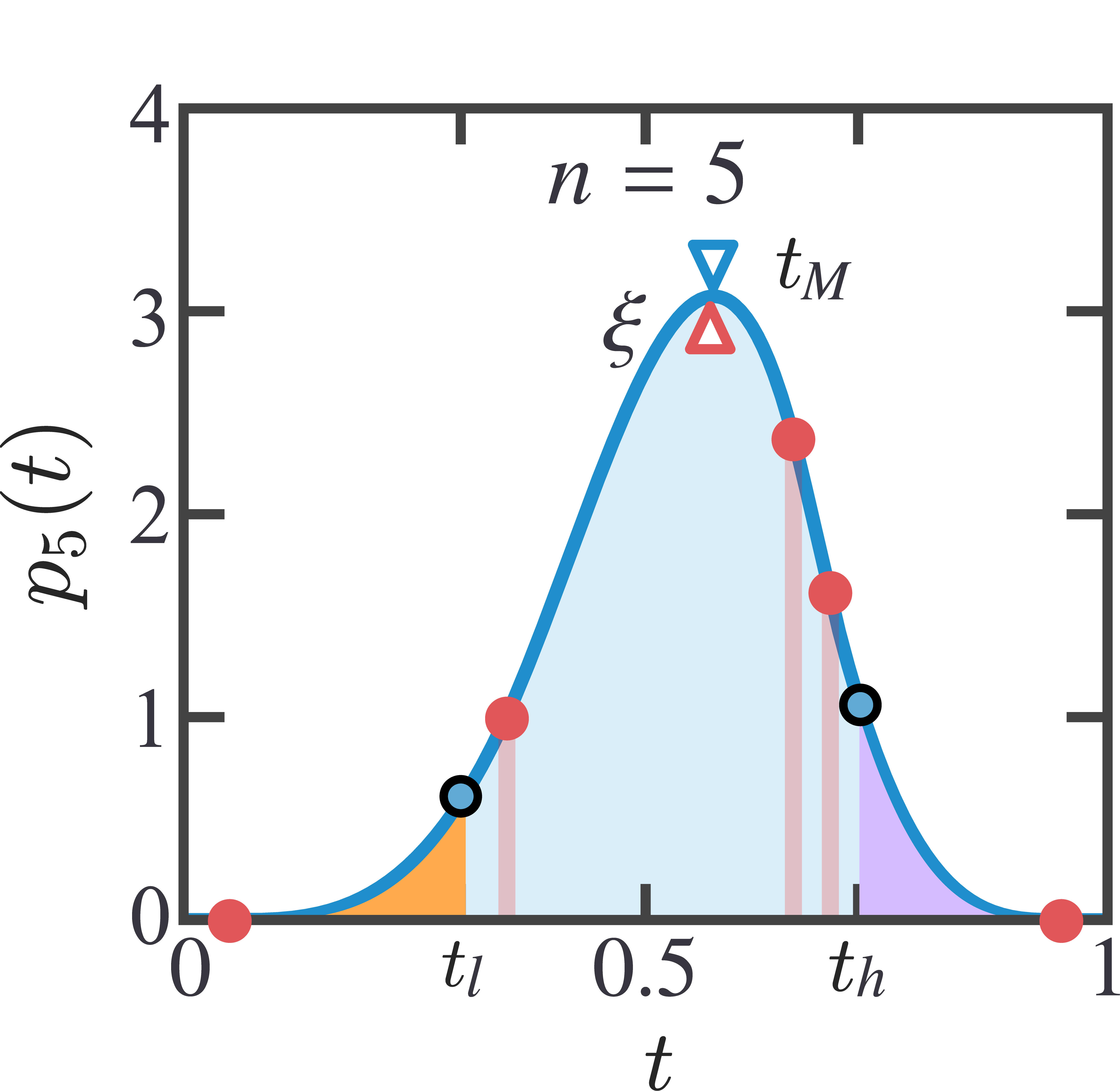}
\caption{Transmissivity distribution $p_n(t)$ for $n=5$ input ports. The blue curve shows the analytical result, with red dots and lines marking the transmission eigenvalues. The blue triangle indicates the mode $t_M$ (peak position), which is close to the Greville abscissa $\xi$ (red triangle). Orange- and purple-shaded regions indicate tail probabilities for extreme events.}
\label{fig:Bspline-shape}
\end{figure}

\textit{Shape properties}---Having examined the cases for $n=2$, $3$, and $4$ in Fig.~\ref{fig:system}(b-d), we now consider $n\geq 5$. Figure~\ref{fig:Bspline-shape} shows $p_5(t;T_5)$ with $\bm{\lambda}(T_5) = (0.05, 0.3, 0.6, 0.7, 0.95)$. As expected, $p_5(t;T_5)$ is a piecewise cubic function with continuous second derivatives at each knot. Both $p_4(t)$ and $p_5(t)$ exhibit a unimodal bell shape with a single peak and two tails—a pattern that holds generally for $n\geq 4$ with nondegenerate $\bm{\lambda}(T)$. (See~\hyperref[em:3]{SM III} for a proof of unimodality.) In general, $p_n(t)$ can be partitioned into three distinct regions:
\begin{equation}
p_n(t) = \begin{cases}   
c_l (t - \lambda_1)^{n-2}, & t \in [\lambda_1, \lambda_2]  \\
\text{central bulk}, & t \in (\lambda_2, \lambda_{n-1}) \\
c_r (\lambda_n - t)^{n-2}, & t \in [\lambda_{n-1}, \lambda_{n}] \\
0, & \text{otherwise}
\end{cases}
\end{equation}

(i) Left tail ($\lambda_1 \leq t \leq \lambda_2$): The density vanishes as $t \to \lambda_1$ following a power law $(t - \lambda_1)^{n-2}$, with coefficient
\begin{equation}\label{eq:bound_cl}
0 < c_l \leq \frac{n-1}{(\lambda_2 - \lambda_{1})^{n-1}}.
\end{equation}
The upper bound is derived from the constraint \begin{equation}\label{eq:bound_cl_derivation}
\int_{\lambda_1}^{\lambda_2} p_n(t) \, \mathrm{d}t \leq 1.
\end{equation}
It is achieved when $\lambda_2 = \lambda_3 = \cdots = \lambda_n$.

(ii) Central bulk ($\lambda_2 < t < \lambda_{n-1}$): The density is log-concave~\cite{gallay2012a} and attains its unique maximum at the mode $t_{M}$, which approximates the Greville abscissa $\xi$~\cite{barnhill1974,piegl1997,farin2002}:
\begin{equation}\label{eq:node}
t_M \approx \xi \coloneqq \frac{1}{n-2}(\lambda_2 + \lambda_3 + \cdots + \lambda_{n-1}).
\end{equation}

(iii) Right tail ($\lambda_{n-1} \leq t \leq \lambda_n$): The density vanishes as $t \to \lambda_n$ following $(\lambda_n - t)^{n-2}$, with coefficient
\begin{equation}\label{eq:bound_cr}
0 < c_r \leq \frac{n-1}{(\lambda_n - \lambda_{n-1})^{n-1}}.
\end{equation}
The upper bound is achieved when $\lambda_1 = \cdots = \lambda_{n-1}$.

This analysis reveals the distinct roles of different eigenvalues: extremal eigenvalues $\lambda_1$ and $\lambda_n$ determine the support and tail behavior, while interior eigenvalues $\lambda_2, \ldots, \lambda_{n-1}$ shape the central bulk and peak position. The second-smallest and second-largest eigenvalues $\lambda_2$ and $\lambda_{n-1}$ play a particularly crucial role, simultaneously defining the central bulk boundaries and controlling tail decay rates through Eqs.~(\ref{eq:bound_cl}) and (\ref{eq:bound_cr}).

The unimodal profile of $p_n(t)$ contrasts sharply with the bimodal eigenvalue distribution in chaotic systems, indicating that extreme transmission events are rare~\cite{mello1988macroscopic,goetschy2013filtering,gerardin2014full}. Using Eq.~(\ref{eq:bound_cl}), we bound the probability that $t$ falls below a threshold $t_l \in [\lambda_1, \lambda_2]$:
\begin{equation}\label{eq:probability_low_threshold}
P(t < t_l) \coloneqq \int_{\lambda_1}^{t_l} p_n(t)\, \mathrm{d}t \leq \left( \frac{t_l - \lambda_1}{\lambda_2 - \lambda_1} \right)^{n-1}.
\end{equation}
Similarly, using Eq.~(\ref{eq:bound_cr}), we bound the probability that $t$ exceeds a threshold $t_h \in [\lambda_{n-1}, \lambda_n]$:
\begin{equation}\label{eq:probability_high_threshold}
P(t > t_h) \coloneqq \int_{t_h}^{\lambda_n} p_n(t)\, \mathrm{d}t \leq \left( \frac{\lambda_n - t_h}{\lambda_n - \lambda_{n-1}} \right)^{n-1}.
\end{equation}

\begin{figure}[hbtp]
\centering
\includegraphics[width = 0.38\textwidth]{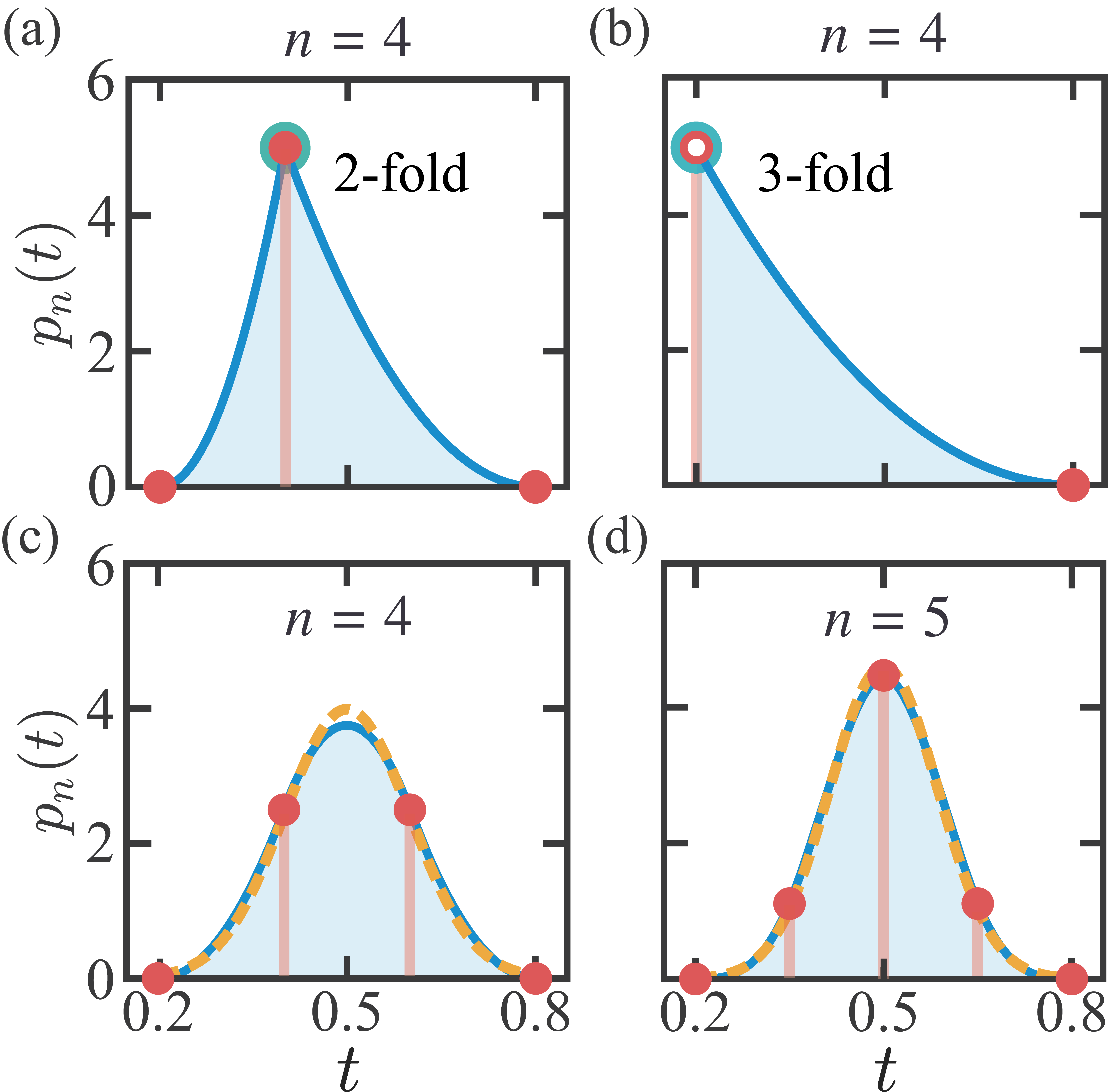}
\caption{Effects of eigenvalue degeneracy and asymptotic behavior. (a,b) Transmissivity distribution $p_4(t)$ with twofold and threefold eigenvalue degeneracy exhibits reduced smoothness at the degeneracy point. (c,d) Distributions $p_4(t)$ and $p_5(t)$ compared with their Gaussian approximations (orange dashed curves) demonstrate the central limit theorem.}
\label{fig:central_limit_theorem}
\end{figure}

\textit{Effects of eigenvalue degeneracy}---We have considered nondegenerate cases where all eigenvalues are distinct. We now extend our theory to degenerate cases where some eigenvalues coalesce. We present two degenerate cases for $n=4$ and compare with the nondegenerate case in Fig.~\ref{fig:system}(d). Figure~\ref{fig:central_limit_theorem}(a) shows $p_4(t;T'_4)$ with $\bm{\lambda}(T'_4) = (0.2,0.4,0.4,0.8)$, exhibiting a sharp corner at the twofold degenerate point $t = 0.4$. The function approaches this maximum quadratically from both sides. Figure~\ref{fig:central_limit_theorem}(b) shows $p_4(t;T''_4)$ with $\bm{\lambda}(T''_4) = (0.2,0.2,0.2,0.8)$, where the threefold degeneracy at $t = 0.2$ creates a boundary maximum with quadratic decay to the right.

These examples illustrate a general principle: each additional eigenvalue coalescing at $\lambda_d$ reduces the smoothness of $p_n(t)$ by one derivative order at $t=\lambda_d$~\cite{curry1966}. Specifically, for nondegenerate eigenvalues, $p_n(t)$ possesses $(n-3)$ continuous derivatives at each eigenvalue when $n\geq 3$. An $l$-fold degeneracy at $\lambda_d$ reduces this to $(n-2-l)$ continuous derivatives. Three important cases emerge: (i) when $l = n-2$, $\lambda_d$ becomes a corner point where $p_n(t)$ attains its global maximum with distinct left and right derivatives [Fig.~\ref{fig:central_limit_theorem}(a)]; (ii) when $l = n-1$, $\lambda_d$ marks a boundary maximum with a one-sided discontinuity [Fig.~\ref{fig:central_limit_theorem}(b)]; (iii) when $l = n$, $p_n(t)$ collapses to a Dirac delta function at $\lambda_d$. The shape of $p_n(t)$ can thus be tailored by strategically merging eigenvalues.

\textit{Moments}---The moments of $p_n(t)$ provide key insights into its shape~\cite{bertsekas2008} and are completely determined by $\bm{\lambda}(T)$. The mean of $p_{n}(t)$ equals the mean of $\bm{\lambda}(T)$~\cite{curry1966}:
\begin{equation}
\mu \coloneqq \int_{-\infty}^{\infty} t \, p_{n}(t) \, \mathrm{d} t = \frac{1}{n}\sum_{i=1}^n \lambda_i.
\end{equation}
The variance of $p_{n}(t)$ equals the variance of $\bm{\lambda}(T)$ divided by $(n+1)$~\cite{curry1966}:
\begin{equation}
\sigma^2 \coloneqq \int_{-\infty}^{\infty} (t-\mu)^2 \, p_{n}(t) \, \mathrm{d} t = \frac{1}{(n+1)n^{2}} \sum_{i>j} (\lambda_i - \lambda_j)^2.
\end{equation}
More generally, the $k$-th raw moment is~\cite{gallay2012a}
\begin{equation}\label{eq:mu_k}
\mu_k \coloneqq \int_{-\infty}^{\infty} t^k \, p_{n}(t) \, \mathrm{d} t = \frac{k! (n-1)!}{(n+k-1)!}\sum_{|\beta|=k} \lambda^\beta,
\end{equation}
where $\lambda^\beta = \lambda_1^{\beta_1} \cdots \lambda_n^{\beta_n}$ and the sum runs over all multi-indices $\beta \in \mathbb{N}^n$ with $|\beta| = \beta_1 + \cdots + \beta_n = k$~\cite{gallay2012a}. Since the collection of all raw moments uniquely determines a probability distribution on a bounded interval~\cite{hausdorff1921}, this confirms that $\bm{\lambda}(T)$ completely determines $p_n(t)$.

We utilize the moments to bound tail probabilities of extreme events. For $n\geq 3$, $p_n(t)$ is unimodal and satisfies a refined Chebyshev inequality known as the one-sided Vysochanskii-Petunin inequality~\cite{mercadier2021one} (see~\hyperref[em:4]{SM IV}):
\begin{align}
P(t < t_a) &\leq \frac{4}{9}\dfrac{\sigma^2}{\sigma^2 + (\mu-t_a)^2}, \;  \text{for\;} t_a < \mu - \sqrt{\frac{5}{3}} \sigma;  \label{eq:Cantelli-low}\\
P(t > t_b) &\leq \frac{4}{9}\dfrac{\sigma^2}{\sigma^2 + (t_b - \mu)^2}, \;  \text{for\;} t_b > \mu + \sqrt{\frac{5}{3}} \sigma.  \label{eq:Cantelli-high}   
\end{align}
These inequalities imply a simple one-sided $3\sigma$ rule: $P(\mu - t > 3\sigma) \leq 0.0445$ and $P(t-\mu > 3\sigma) \leq 0.0445$. Applying inequality~(\ref{eq:Cantelli-high}) to the distribution in Fig.~\ref{fig:Bspline-shape} yields $P(t >0.75)\leq 0.105$, whereas the exact value is $0.031$. These results confirm the rarity of extreme events.

\textit{Asymptotic behaviors}---We now examine the asymptotic behavior of $p_n(t)$ for large $n$. Figures~\ref{fig:system}(d) and~\ref{fig:Bspline-shape} suggest that $p_n(t)$ can resemble a Gaussian distribution even for moderate values such as $n=4$ or $5$. This observation is explained by a central limit theorem for B-splines: consider a sequence of transmission matrices $\{T_n\}$, where $T_n$ is of size $n\times n$ with $\bm{\lambda}(T_n) = (\lambda_1^{(n)}, \ldots, \lambda_n^{(n)})$. If the sequence $\{\bm{\lambda}(T_n)\}$ satisfies
\begin{align}
  &\lim_{n \to \infty} \frac{\log n }{\sqrt{n}} \lambda^{(n)}_n = 0, \label{eq:CLT-condition-1} \\
  &\lim_{n \to \infty} \frac{1}{n}\sum_{i=1}^n \lambda_i^{(n)} = \mu, \label{eq:CLT-condition-2}\\
  &\lim_{n \to \infty} \frac{1}{n^{2}} \sum_{i>j} (\lambda_i^{(n)} - \lambda_j^{(n)})^2 = s^2, \label{eq:CLT-condition-3}
 \end{align}
then $p_{n}(t; T_n)$ converges weakly to the normal distribution $\mathcal{N}(\mu, \sigma^2)$ with $\sigma^2 = s^2 / (n+1)$ as $n \to \infty$~\cite{gallay2012a}.

For passive systems where $0 \leq \lambda_1^{(n)} \leq \cdots \leq \lambda_n^{(n)} \leq 1$, condition~(\ref{eq:CLT-condition-1}) is automatically satisfied. Under the mild assumptions~(\ref{eq:CLT-condition-2}) and~(\ref{eq:CLT-condition-3}), $p_n(t)$ approaches a Gaussian distribution. Furthermore, the variance of the limiting Gaussian satisfies
\begin{equation}
    \sigma^2 \leq \frac{1}{4(n+1)},
\end{equation}
with equality when half the eigenvalues equal $0$ and half equal $1$. The probability distribution thus concentrates increasingly around its mean $\mu$ as $n$ grows.

As illustrations, Figure~\ref{fig:central_limit_theorem}(c) shows $p_4(t;\tilde{T}_4)$ with $\bm{\lambda}(\tilde{T}_4) = (0.2,0.4,0.6,0.8)$ alongside its Gaussian fit $\mathcal{N}(0.5, 0.01)$. Figure~\ref{fig:central_limit_theorem}(d) shows $p_5(t;\tilde{T}_5)$ with $\bm{\lambda}(\tilde{T}_5) = (0.2,0.35,0.5,0.65,0.8)$ and its Gaussian fit $\mathcal{N}(0.5, 0.0075)$. The Gaussian approximation is already quite accurate for $n=5$~\cite{unser1992}.

This central limit theorem has significant physical implications. While $p_n(t;T)$ is determined by the transmission eigenvalues $\bm{\lambda}(T)$, in the large-$n$ limit the distribution converges to a Gaussian characterized solely by the mean and variance of $\bm{\lambda}(T)$. All higher-order information becomes irrelevant. Crucially, the detailed shape of the transmission eigenvalue distribution $p[\bm{\lambda}(T)]$---whether bimodal as in Figs.~\ref{fig:intro}(b,c) or Bernoulli as in Figs.~\ref{fig:intro}(d,e)---affects only the position and width of the resulting Gaussian profile. This resolves the apparent paradox in Fig.~\ref{fig:intro}.

\begin{figure}[htbp]
\centering
\includegraphics[width = 0.48\textwidth]{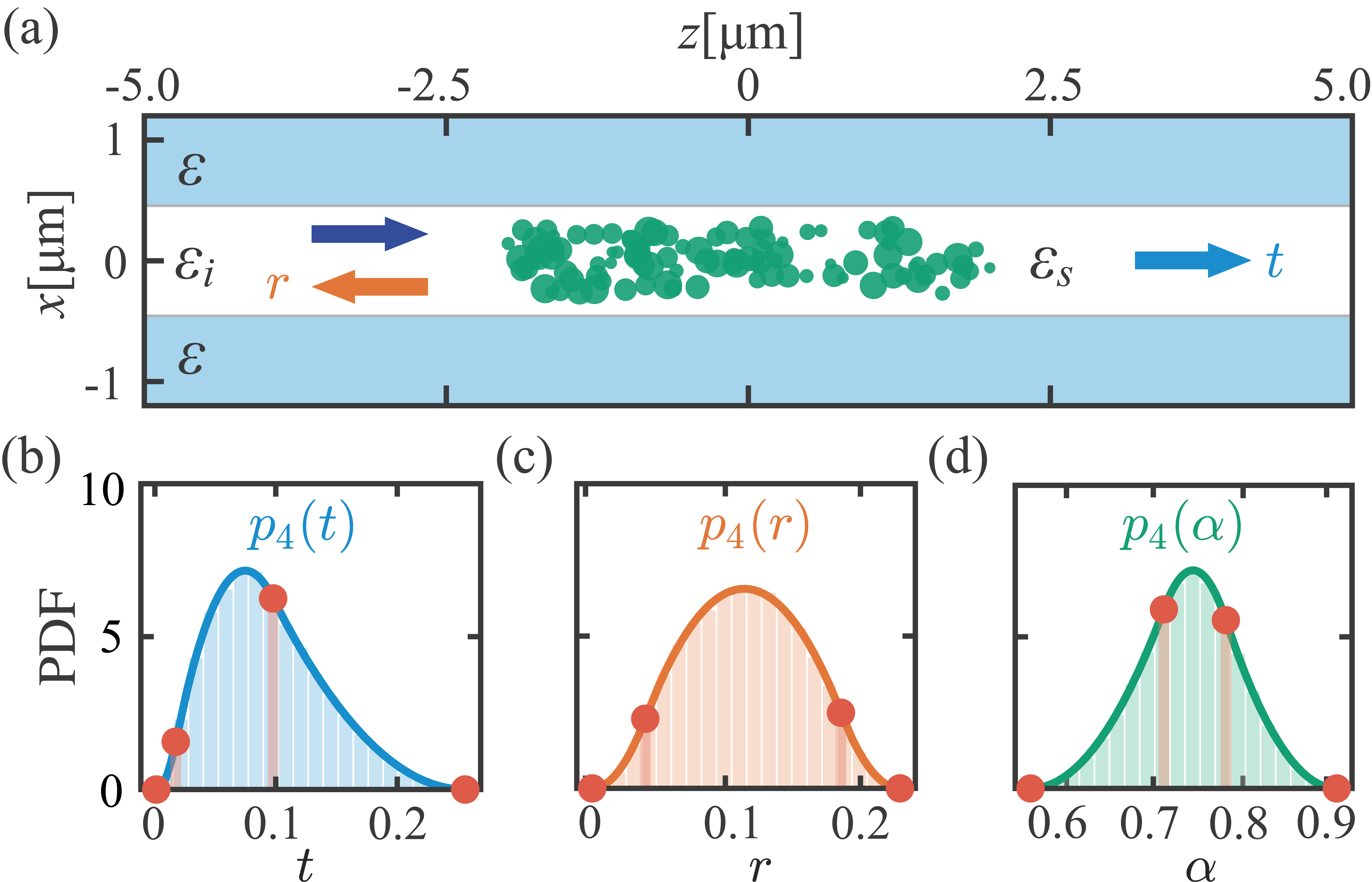}
\caption{Numerical demonstration of the theory. (a) A slab waveguide with permittivities $\varepsilon_i = 12.1$ (core), $\varepsilon = 2.1$ (cladding), and $\varepsilon_s = 2.0+0.86i$ (lossy scatterers), supporting $n=4$ TE modes. (b-d) Probability distributions for transmissivity $p_4(t)$, reflectivity $p_4(r)$, and absorptivity $p_4(\alpha)$. Histograms show Monte Carlo results from $10^5$ random coherent inputs, and solid curves show theoretical B-spline predictions.}
\label{fig:real-system}
\end{figure}

\textit{Generalization and demonstration}---Our theory extends naturally beyond transmission to encompass a broad class of transport observables. For any measurement where a coherent input wave $\bm{a}$ yields an observable
\begin{equation}
    o[\bm{a}] = \bm{a}^\dagger O \bm{a},
\end{equation}
with $O$ a Hermitian operator, the probability distribution of $o[\bm{a}]$ for a random coherent input $\bm{a}$ is a fundamental B-spline with knots given by the eigenvalues $\bm{\lambda}(O)$.

For example, consider again the linear system in Fig.~\ref{fig:system}(a) with the block scattering matrix in Eq.~(\ref{eq:block-S-matrix}). The reflectivity and absorptivity are given by~\cite{guo2024passivity,guo2025a}
\begin{equation}
r[\bm{a}] = \bm{a}^{\dagger} R\bm{a}, \quad \alpha[\bm{a}] = \bm{a}^{\dagger} A \bm{a},
\end{equation}
where $R=\rho^\dagger \rho$ is the reflectance matrix~\cite{guo2024unitary} and $A = I-\tau^\dagger \tau -\rho^\dagger \rho$ is the absorptivity matrix~\cite{guo2023a}. Our probability theory applies directly to these observables.

As an illustration, we consider a disordered dielectric slab waveguide [Fig.~\ref{fig:real-system}(a)] with a silicon core ($\varepsilon_{i} = 12.1$) and silica cladding ($\varepsilon = 2.1$). The core contains $100$ randomly positioned lossy silica cylinders ($\varepsilon_s=2.0+0.86i$). The waveguide supports $n=4$ TE modes at $\lambda_0 = \SI{1.55}{\micro\meter}$. Using the FDTD method~\cite{Hughes2021A}, we calculate $t[\bm{a}]$, $r[\bm{a}]$, and $\alpha[\bm{a}]$ for $10^5$ random coherent inputs $\bm{a}$. Figures~\ref{fig:real-system}(b-d) show the resulting histograms alongside the theoretical B-spline distributions $p_4(t)$, $p_4(r)$, and $p_4(\alpha)$ determined by the eigenvalues of the corresponding matrices $T$, $R$, and $A$. The theoretical curves agree well with the simulation results in all three cases. (See~\hyperref[em:5]{SM V} for computational details and $n=3$ results.)

\textit{Final remarks and conclusion}---We make four concluding remarks. First, our results apply to both classical and quantum waves, including optical, acoustic, and electronic systems. Second, while we focus on transport measurements, our probability framework extends naturally to other types of observables. Third, our theoretical predictions---including the probability distribution and its moments---are directly measurable using current wavefront shaping techniques. Fourth, our bounds on tail probabilities [Eqs.~(\ref{eq:probability_low_threshold}, \ref{eq:probability_high_threshold}, \ref{eq:Cantelli-low}, \ref{eq:Cantelli-high})] can be used to evaluate the likelihood of extreme phenomena such as open channels, coherent perfect absorption~\cite{dai2018topologically,guo2024unitary,guo2024passivity}, and reflectionless scattering modes~\cite{sweeney2020theory,sol2023reflectionless,jiang2024coherent}.

In conclusion, we have established that the probability distribution of transmissivity for random coherent waves through arbitrary media follows a fundamental B-spline with knots determined by the transmission eigenvalues. We reveal that the transmissivity distribution converges to a Gaussian in the large-$n$ limit, with its mean and variance determined solely by the first two moments of the eigenvalue distribution. This result resolves the apparent paradox between bimodal transmission eigenvalue distribution and unimodal transmissivity distribution. Our theoretical framework extends to other experimental observables and provides a rigorous foundation for understanding wave transport statistics in complex media.

\begin{acknowledgments}
C.G. is supported by the Jack Kilby/Texas Instruments Endowed Faculty Fellowship.
\end{acknowledgments}
\bibliography{main}

\clearpage
\newpage

\appendix
\setcounter{equation}{0}
\renewcommand{\theequation}{E.\arabic{equation}}
\setcounter{figure}{0}
\renewcommand{\thefigure}{S\arabic{figure}}
\onecolumngrid
\section{Supplementary Material}
\subsection{SM I. Expressions of $T_3$ and $T_4$}\label{em:1}

In this section, we provide explicit expressions of $T_3$ and $T_4$ used in Figs.~\ref{fig:system}(c) and \ref{fig:system}(d):

\begin{center}
\begin{minipage}{\textwidth}

\centering
\begin{equation}
    T_3=\begin{pmatrix}
        0.569 & 0.095 + 0.039j & 0.102 - 0.072j\\0.095 - 0.039 j & 0.647 & 0.052 - 0.157 j\\ 0.101 + 0.072 j& 0.052 + 0.157 j & 0.284
    \end{pmatrix},
\end{equation}
\begin{equation}
    T_4=\begin{pmatrix}
    0.520 & -0.078 + 0.121 j & -0.071 - 0.007 j & 
   0.167 + 0.085 j\\
  -0.078 - 0.122j& 
   0.633& -0.166 + 0.057 j& -0.030 + 0.100 j\\
  -0.071 + 0.007 j & -0.166 - 0.057 j& 0.435&
   0.023 + 0.006 j\\
  0.167 - 0.0985 j& -0.030 - 0.100 j& 
   0.023 - 0.006 j& 0.513
    \end{pmatrix}.
\end{equation}

\end{minipage}
\end{center}

\vspace{12pt}
\twocolumngrid
\subsection{SM II. Proof of Eq.~(\ref{eq:main_result})}
\label{em:2}

In this section, we provide a proof of Eq.~(\ref{eq:main_result}).

\begin{proof}
This proof follows from Proposition 8.2 in Ref.~\cite{olshanski1996}. See also Proposition 3.1 in Ref.~\cite{gallay2012a}. 

Our goal is to determine $p_n(t;T)$, the probability distribution of $t = \bm{a}^\dagger T \bm{a}$ where $\bm{a}$ is drawn from a uniform distribution (Haar measure) on the complex sphere $\mathcal{S}^{2n-1}$. We begin by diagonalizing the Hermitian matrix $T = V^\dagger \Lambda V$, where $V$ is unitary and $\Lambda = \operatorname{diag}(\lambda_1, \ldots, \lambda_n)$. With the change of variables $\bm{a}' = V\bm{a}$, we can rewrite the transmissivity as $t = \bm{a}'^\dagger \Lambda \bm{a}'$. Since the Haar measure is invariant under unitary transformations, $\bm{a}'$ is also uniformly distributed on $\mathcal{S}^{2n-1}$. This yields
\begin{equation}
t = \sum_{i=1}^{n} \lambda_i |a'_i|^2.
\end{equation}

This expression motivates us to consider the map
\begin{equation}
\bm{a}' = (a'_1, \ldots, a'_n) \mapsto \bm{s} = (|a'_1|^2, \ldots, |a'_n|^2),
\end{equation}
which maps the sphere $\mathcal{S}^{2n-1}$ to the standard $(n-1)$-simplex $\sigma_{n-1}= \{(s_1, \ldots, s_n) \in \mathbb{R}^n \mid 0 \le s_1, \ldots, s_n \le 1,\, s_1 + \cdots + s_n = 1\}$. Under this map, the uniform measure on $\mathcal{S}^{2n-1}$ pushes forward to the uniform measure on $\sigma_{n-1}$~\cite{olshanski1996}. We have thus transformed the original problem into determining the probability distribution of $t = \bm{\lambda} \cdot \bm{s}$, where $\bm{s}$ is a random vector uniformly distributed over the simplex $\sigma_{n-1}$. This distribution is precisely the fundamental B-spline in Eq.~(\ref{eq:main_result}), as established by the following geometric interpretation of the fundamental B-spline (see Theorem 2 of Ref.~\cite{curry1966} and Ref.~\cite{olshanski1996} \S8):

\begin{theorem*}[Curry and Schoenberg, 1966~\cite{curry1966}]
The fundamental $B$-spline $M_{n-1}(t;\lambda_1, \ldots, \lambda_n)$ is the linear density function obtained by orthogonal projection onto the $t$-axis of an $(n-1)$-dimensional simplex $\sigma_{n-1}$ with unit volume, positioned such that its $n$ vertices project orthogonally onto the points $\lambda_1, \lambda_2, \ldots, \lambda_n$ on the $t$-axis.
\end{theorem*}

This completes the proof of Eq.~(\ref{eq:main_result}).
\end{proof} 

\subsection{SM III. Proof of Unimodality}
\label{em:3}

In this section, we prove $p_n(t)$ is unimodal when $n\geq 3$.

\begin{proof}
For simplicity, we assume non-degenerate eigenvalues. Degenerate cases can be proven by a continuity argument. 
According to our central result Eq.~(\ref{eq:main_result}),
\begin{equation}\label{eq:main_result_reproduce}
    p_n(t;T) = M_{n-1}[t;\bm{\lambda}(T)].
\end{equation}
When $n=3$, $p_n(t)$ is a triangular distribution and thus unimodal. For $n\geq 4$, we invoke the following theorem:
\begin{proposition}[Curry and Schoenberg, 1966~\cite{curry1966}]
Consider a fundamental B-spline $M_{n-1}(t)$ ($n\geq 4$) with non-degenerate knots $\lambda_1 < \lambda_2 < \cdots < \lambda_n$. Then its $\nu$-th order derivative $M_{n-1}^{(\nu)}(t)$ ($\nu = 0, \ldots, n-3$) has exactly $\nu$ distinct simple zeros in the interval $(\lambda_1,\lambda_n)$. 
\end{proposition}
Setting $\nu =1$, we see that the first derivative of $p_n(t)=M_{n-1}(t)$ has exactly one simple zero for $n\geq 4$. This establishes the unimodality of $p_n(t)$ when $n\geq 3$. 
\end{proof}

\subsection{SM IV. Chebyshev-type Inequalities}
\label{em:4}

In this section, we review Chebyshev-type inequalities~\cite{savage1961}, including the one-sided Vysochanskii--Petunin inequalities used to obtain Eqs.~(\ref{eq:Cantelli-low}) and (\ref{eq:Cantelli-high}).

Let $X$ be a random variable with mean $\mu$ and variance $\sigma^2$. Chebyshev's inequality states
\begin{equation}
    P(|X - \mu| \geq c) \leq \frac{\sigma^2}{c^2}, \quad \text{for all $c>0$.}
\end{equation}
This inequality has a one-sided refinement known as Cantelli's inequality~\cite{cantelli1929}:
\begin{equation}
P(X - \mu \geq c) \leq \frac{\sigma^2}{\sigma^2 + c^2}, \quad \text{for all $c>0$.}
\end{equation}

If $X$ is further assumed to have a unimodal continuous distribution, these inequalities can be refined. Chebyshev's inequality can be sharpened to the Vysochanskii--Petunin inequality~\cite{pukelsheim1994three}:
\begin{equation}
P(|X - \mu | \geq c) \leq
    \begin{cases}
        \dfrac{4}{9}\dfrac{\sigma^2}{c^2}, & c \geq \sqrt{\dfrac{8}{3}} \sigma , \\[12pt]
        \dfrac{4}{3}\dfrac{\sigma^2}{c^2} - \dfrac{1}{3}, & 0 < c \leq \sqrt{\dfrac{8}{3}} \sigma.
    \end{cases}
\end{equation}
Similarly, Cantelli's inequality can be refined to the one-sided Vysochanskii--Petunin inequality~\cite{mercadier2021one}:
\begin{equation}
P(X - \mu \geq c) \leq 
    \begin{cases}
        \dfrac{4}{9} \dfrac{\sigma^2}{\sigma^2+c^2}, & c \geq \sqrt{\dfrac{5}{3}} \sigma , \\[12pt]
        \dfrac{4}{3} \dfrac{\sigma^2}{\sigma^2+c^2} - \dfrac{1}{3}, & 0 < c \leq \sqrt{\dfrac{5}{3}} \sigma.
    \end{cases}
\end{equation}

\subsection{SM V. Simulation Details}
\label{em:5}

\begin{figure}[hbtp]
\centering
\includegraphics[width = 0.48\textwidth]{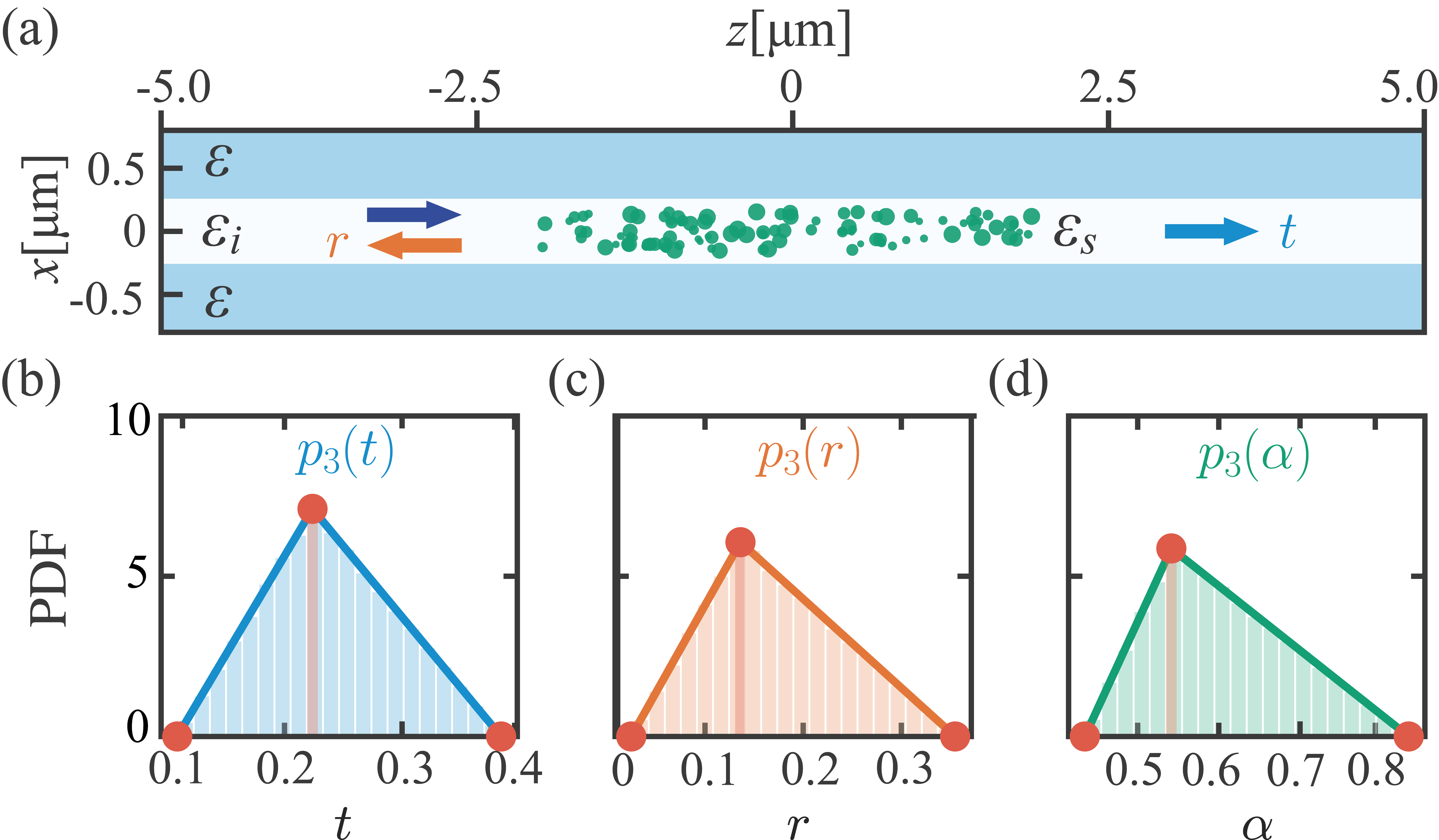}
\caption{Another numerical demonstration. (a) A slab waveguide with permittivities $\varepsilon_i=12.1$ (core), $\varepsilon=2.1$ (cladding), and $\varepsilon_s=2.0+0.86i$ (lossy scatterers), supporting $n=3$ TE modes. (b--d) Probability distributions for transmissivity $p_3(t)$, reflectivity $p_3(r)$, and absorptivity $p_3(\alpha)$. Histograms show Monte Carlo results from $10^5$ random coherent inputs, and solid curves show theoretical B-spline predictions.}
\label{fig:real-system-2}
\end{figure}

In this section, we provide computational details for the numerical demonstration in Fig.~\ref{fig:real-system}. We also present additional simulation results for a disordered waveguide supporting $n=3$ TE modes (see Fig.~\ref{fig:real-system-2}).

In our numerical demonstration, we consider a slab waveguide comprising a silicon core ($\varepsilon_{i} = 12.1$) embedded in silica cladding ($\varepsilon = 2.1$). The waveguide has width $w$ in the $x$ direction and is uniform in the $y$ direction. Light propagates in the $z$ direction with vacuum wavelength $\lambda_0 = \SI{1.55}{\micro\meter}$; the electric field is polarized along $y$ (TE polarization). The uniform waveguide supports $n$ eigenmodes at $\lambda_0$, with $n$ depending on $w$. In Fig.~\ref{fig:real-system}, we set $w=\lambda_0/1.7$ and obtain $n=4$ modes. In Fig.~\ref{fig:real-system-2}, we set $w=\lambda_0/3$ and obtain $n=3$ modes.

To introduce disorder and loss, we randomly place 100 cylindrical scatterers within the waveguide. These scatterers consist of lossy silica with complex relative permittivity $\varepsilon_s=2.0+0.86i$. Their centers are distributed in the region $-\SI{2}{\micro\meter} \leq z \leq \SI{2}{\micro\meter}$ and $-0.3w \leq x \leq 0.3w$. Their diameters are uniformly distributed over $[0.10w, 0.27w]$. We place perfectly matched layers outside the cladding to absorb all leaky radiation due to scattering. 

We simulate the light propagation in the waveguide using Tidy3D~\cite{Hughes2021A}, which implements the finite-difference time-domain method. We excite and launch the guided eigenmodes from the left port at $z=-\SI{4}{\micro\meter}$ and compute the transmitted and reflected fields at $z=\SI{4}{\micro\meter}$ and $z=-\SI{4.38}{\micro\meter}$, respectively. We then decompose these output fields using the guided-mode bases at the left and right ports. By exciting each guided eigenmode in turn, we obtain the field transmission and reflection blocks, $\tau$ and $\rho$, of the scattering matrix $S$ in Eq.~(\ref{eq:block-S-matrix}). We then calculate the power operators $T = \tau^\dagger \tau$, $R = \rho^\dagger \rho$, and $A = I - T - R$. Using the Monte Carlo method, we calculate $t[\bm{a}]=\bm{a}^{\dagger} T\bm{a}$, $r[\bm{a}]=\bm{a}^{\dagger} R\bm{a}$, and $\alpha[\bm{a}]=\bm{a}^{\dagger} A\bm{a}$ for $10^5$ random coherent inputs $\bm{a}$. The results are summarized in Figs.~\ref{fig:real-system}(b--d) and \ref{fig:real-system-2}(b--d). The theoretical curves agree well with the simulation results for both $n=3$ and $n=4$. These results demonstrate the validity of our theory. 

\end{document}